\documentclass[aip, amsmath,amssymb,reprint]{revtex4-1}
\usepackage{graphicx}% Include figure files
\usepackage{dcolumn}% Align table columns on decimal point
\usepackage{bm}% bold math
\usepackage{caption}
\usepackage{mathtools}
\usepackage{subcaption}

\begin{document}

\title{
Sheet model description of spatio-temporal evolution of upper-hybrid oscillations in an inhomogeneous magnetic field }
\author{Nidhi Rathee}
\email[]{ratheenidhi822@gmail.com}
\altaffiliation[]{Present address: School of Mathematics and Physics, Queen’s University Belfast, University Road, Belfast BT7 1NN, UK}
\affiliation{Institute for Plasma Research, Gandhinagar, Gujarat, 382428, India}
\affiliation{Homi Bhabha National Institute, Training School Complex, Mumbai, 400094, India}
\author{Someswar Dutta}
\affiliation{Institute for Plasma Research, Gandhinagar, Gujarat, 382428, India}
\author{R. Srinivasan}
\affiliation{Institute for Plasma Research, Gandhinagar, Gujarat, 382428, India}
\affiliation{Homi Bhabha National Institute, Training School Complex, Mumbai, 400094, India}
\author{Sudip Sengupta}
\affiliation{Institute for Plasma Research, Gandhinagar, Gujarat, 382428, India}
\affiliation{Homi Bhabha National Institute, Training School Complex, Mumbai, 400094, India}

\date{\today}

\begin{abstract}
Spatio-temporal evolution of large amplitude upper hybrid oscillations in a cold homogeneous plasma in the presence of an inhomogeneous magnetic field is studied analytically and numerically using the Dawson sheet model\cite{dawson_sheet}. It is observed that the inhomogeneity in magnetic field which causes the upper hybrid frequency to acquire a spatial dependence, results in phase mixing and subsequent breaking of the upper hybrid oscillations at arbitrarily low amplitudes. This result is in sharp contrast to the usual upper hybrid oscillations in a homogeneous magnetic field  where the oscillations break within a fraction of a period when the amplitude exceeds a certain critical value\cite{Davidson72}. 
Our perturbative calculations show that the phase mixing (wave breaking) time scales inversely with the amplitude of magnetic field inhomogeneity 
($\Delta$) and amplitude of imposed density perturbation ($\delta$),  and scales directly with the ratio of magnetic field inhomogeneity scale length to imposed density perturbation scale length ($(\alpha/k_L)^{-1}$ ) as  $\omega_{pe}\tau_{mix} \sim \left( 1+\beta^2 \right) ^{3/2}k_L/(\beta^2\delta\Delta\alpha)$, where $\beta$ is the ratio of electron cyclotron frequency to electron plasma frequency. Further phase mixing time measured in simulations, performed using a 1-1/2 D code based on Dawson sheet model\cite{dawson_sheet}, shows good agreement with the above mentioned scaling. This result may be of relevance to plasma based particle acceleration experiments in the presence of a transverse inhomogeneous magnetic field.  
%
%The simulation results are in excellent agreement with the recent analytical results presented in ref.\cite{chandanpre}.
% 
\end{abstract}

%\pacs{}% insert suggested PACS numbers in braces on next line

\maketitle 

\section{Introduction}

Spatio-temporal evolution of nonlinear plasma waves and their breaking is a fascinating field of study due to its wide ranging applications, for example in  particle acceleration schemes\cite{tajima79,modena95,malka02,hegelich06,schwoerer06,faure06,matlis06,dieckmann2004},  laser assisted fusion schemes\cite{tabak94,kodama01}, collisionless heating of plasma \cite{koch74,bose15,bauer92,sandhu05}, heating of solar corona \cite{botha00,voitenko05,hasegawa74prl} etc. It is well known that electrons in a cold homogeneous plasma, when linearly or nonlinearly perturbed, exhibit large amplitude oscillations at the usual plasma frequency given by $\omega_{pe}^2  = 4\pi n_0 e^2 / m$ where $\omega_{pe}$ is the electron plasma frequency, $e$ the electron charge, $n_0$ the background ion density and $m$ is the mass of an electron, provided the amplitude of the perturbation is   
below a critical value $\mid \delta n_e/n_0 \mid < 0.5$ (Here ions are assumed to be infinitely massive). This was shown independently by Dawson\citep{dawson59} and Davidson {\it et. al.}\citep{Davidson72} by performing a nonlinear analysis of the Fluid-Maxwell set of equations using Lagrange variable technique. Beyond this critical amplitude, the  waves/oscillations break\cite{dawson59,Davidson72} within a period.

It was further shown by Davidson {\it et. al.}\cite{Davidson72} that in the presence of an uniform transverse external magnetic field, plasma electrons, when linearly or nonlinearly perturbed, exhibit upper hybrid oscillations at a frequency $\omega_{uh0}^2 = \omega_{pe}^2 + \omega_{ce0}^2$, where $\omega_{ce0} = eB_0/mc $, $B_0$ being the external magnetic field and $c$ is speed of light in vacuum, provided the amplitude of perturbation is below a critical limit given by $\mid \delta n_e/n_0 \mid < ( 1 + \omega_{ce0}^2/\omega_{pe}^2) / 2$ (assuming there is no initial shear in the electron fluid velocity). As before, beyond this critical amplitude the oscillations/waves break within a period.
 
The effect of spatial inhomogeneity in the external magnetic field on upper hybrid oscillations was analytically studied by Maity {\it et. al.}\citep{chandanpre}. Following an analysis similar to that used by Infield {\it et.al.}\cite{infeld89} for electron plasma oscillations in a cold unmagnetized plasma with inhomogeneous density, Maity {\it et. al.}\cite{chandanpre} presented an exact solution to the problem of upper-hybrid oscillations in an inhomogeneous magnetic field  in parametric form.
It was found that, inclusion of inhomogeneity in the external magnetic field inevitably results in breaking of the upper-hybrid oscillations at arbitrarily low amplitudes via the process of phase mixing, a result which is in sharp contrast to the usual upper hybrid oscillations in an uniform external magnetic field, where, as mentioned above, coherent oscillations are sustained below a certain critical amplitude\cite{Davidson72}. 
Physically,  the  magnetic  field  inhomogeneity causes the electron cyclotron frequency to acquire spatial dependency, making the upper-hybrid frequency space dependent %the upper hybrid frequency to acquire a spatial dependence 
( {\it i.e.} $\omega_{uh}^2(x) = \omega_{pe}^2 + \omega_{ce}^2(x)$ where $\omega_{ce}(x) = eB(x)/mc$ ). As a result electron fluid elements located at different spatial positions oscillate  at  different  local  frequency,  resulting in mixing of their phases\cite{dawson59}. Eventually the neighbouring oscillators go out of phase and cross each other, resulting in destruction of the coherent motion {\it i.e} breaking of the wave/oscillation. The electron density exhibits a singularity at this point. This is the well known phenomenon of wave breaking via the process of phase mixing \cite{drake76,sudip11ppcf,infeld89,kaw73,chandanpre,sudip99prl}, and the time taken for the wave to break is known as the phase-mixing time.  This physics of phase mixing process, which depends only on spatial variation of the characteristic frequency, is relevant for both non-relativistic and relativistic plasmas\cite{chandanprl}. 
%
%Evidence of phase mixing of relativistic upper hybrid waves has been seen in PIC simulation of laser pulse absorption [].
%
The problem of electrostatic waves (upper-hybrid waves) propagating perpendicular to an inhomogeneous magnetic field has recently acquired prominence because of its application to particle acceleration experiments\cite{artemyev2015prl}

In this paper, we present a detailed numerical simulation of upper hybrid oscillations in both homogeneous and inhomogeneous magnetic field using a 1-1/2 D code based on the Dawson sheet model\cite{dawson_sheet}. 
For both the cases analytical solutions presented in references\cite{Davidson72, chandanpre} are first reproduced here using the physically appealing Dawson sheet model\cite{dawson_sheet}, and thereafter simulation results are compared with the analytical results. 
In section \ref{sec2}, equations governing the evolution of upper-hybrid oscillations are described in terms of Dawson sheet coordinates. In section \ref{homogeneous}, exact solution describing large amplitude upper hybrid oscillations in a homogeneous magnetic field is presented and  conditions ( inequalities ) for sustained coherent motion is derived in terms of sheet coordinates. 
%{\it These inequalities are equivalent to the inequalities presented in Eq. (52) and (53) of chapter 3 of reference\cite{Davidson72}.}
The upper hybrid oscillations break when these inequalities are violated.
In section \ref{inhomogeneous}, exact solution of the sheet equations describing upper hybrid oscillations in an inhomogeneous magnetic field is presented in parametric form. In order to elucidate the physics described by the exact solution presented in section \ref{inhomogeneous}, in subsection \ref{app_analysis} we perform a perturbative analysis of the same equations and present an explicit expression for phase mixing time, clearly exhibiting its dependence on the amplitude and scale length of both the magnetic field inhomogeneity and imposed density perturbation. 
In section \ref{numerical} we describe our simulations, where we excite upper-hybrid oscillations by perturbing the number density of electron sheets and study the spatio-temporal evolution of these oscillations. Simulation results for the homogeneous case are first compared with the analytical results. Next the affect of spatial inhomogeneity in the external magnetic field on the excited upper hybrid oscillations is described. It is found that upon introduction of a spatial inhomogeneity in the external magnetic field, the excited upper hybrid oscillations break at arbitrary low amplitudes. To explore the mechanism of breaking, 
the time taken by two adjacent sheets to cross over (phase mixing / wave breaking time)\cite{sudip09pre} is measured in simulations. Simulation results showing the dependence of phase mixing time ( wave breaking time ) on the density perturbation amplitude, the magnetic field inhomogeneity scale length and  the magnetic field inhomogeneity amplitude are then presented. 
%{\bf  It is found that the scaling of phase mixing time measured in simulations is in good agreement with its analytical expression presented in subsection \ref{app_analysis}.}
% 
Finally in Sec. \ref{sec4} we summarize our results. 

\section{Governing Equations and Sheet Model description of upper-hybrid oscillations} \label{sec2}
In Eulerian coordinates, in the presence of an external  magnetic field ( taken along $z$ direction ), the space time evolution of upper-hybrid mode under electrostatic approximation is governed by the following equations\cite{Davidson72}: 
\begin{eqnarray}
\frac{\partial n_e}{\partial t} + \frac{\partial}{\partial x}(n_e v_x) & = & 0 \label{continuity}\\
\nonumber \\
\frac{\partial v_x}{\partial t} + v_x \frac{\partial v_x}{\partial x} & = & -E -\beta v_y \label{mom_x}\\ 
\nonumber \\
\frac{\partial v_y}{\partial t} + v_x \frac{\partial v_y}{\partial x} & = & \beta v_x \label{mom_y}\\ 
\nonumber \\
\frac{\partial E}{\partial x} & = & (n_i - n_e) \label{poisson}	
\end{eqnarray}
where Eq.(\ref{continuity}) is the electron continuity equation, Eq.(\ref{mom_x}) and 
Eq.(\ref{mom_y}) are respectively the $x$ and $y$ component of momentum equation and 
Eq.(\ref{poisson}) is the Poisson's equation, with the symbols having their usual meaning. Here ions are assumed to be stationary with uniform density $n_0$, thus providing a uniform static neutralizing background.
In the above equations, we have used the following normalization $t\rightarrow \omega_{pe}t \; , \; x\rightarrow k_L x \; , \; v\rightarrow k_L v/\omega_{pe} \; , \; n_e\rightarrow n_e/n_0 \; , \;  E\rightarrow k_LeE/m\omega_{pe}^2  $
%\; ; \; \alpha\rightarrow \alpha/k$
and $\beta=\omega_{ce}/\omega_{pe}  \; , \; \omega_{ce}=eB_0/mc $. Here $k_L$ is the mode number of the imposed density perturbation.
We now introduce Lagrange coordinates $(x_0,\tau)$ ( Dawson coordinates\cite{dawson59,dawson_sheet} ), which are related to the Euler coordinates $(x,t)$ as
\begin{equation}\label{eqn1}
x =  x_0 + \xi (x_0, \tau) , \quad t =  \tau
\end{equation}
Defining $v_x(x_0,\tau) = \partial \xi / \partial \tau $, Eqs.(\ref{continuity}) -  (\ref{poisson}), respectively transform as 
\begin{eqnarray}
n_e(x_0,\tau) & = & \left[ 1+\frac{\partial \xi}{\partial x_0}\right] ^{-1} \label{continuity_L}\\
\nonumber \\
\ddot{\xi} & = & -\xi - \beta v_y \label{mom_xL}\\
\nonumber \\
\dot{v}_y & = &\beta\dot{\xi} \label{mom_yL}\\
\nonumber \\
E(x_0,\tau) & = & \xi  \label{poissonL}
\end{eqnarray}
where 'dot' represents derivative w.r.t Lagrange time $\tau$.

The above equations (Eq.(\ref{continuity_L})- Eq.(\ref{poissonL})) are amenable to a neat geometrical interpretation, as shown in the Fig.(1). Here electrons are assumed to be infinite sheets of charge embedded in a cold immobile positive ion background\cite{dawson59}. $x_0$ (shown as solid vertical line) and $\xi(x_0,\tau)$ (shown as dotted vertical line), are respectively the equilibrium position and displacement from the equilibrium position of an electron sheet. Equilibrium position of an electron sheet coincides with the position of an ion. The external magnetic field is taken along the $z$ direction and in Fig.(1) it is shown to be inhomogeneous along $x$ direction. Counting the number of sheets per unit length at any time $\tau$, immediately leads to Eq.(\ref{continuity_L})\cite{birdsall85}, whereas the displacement of a sheet from its equilibrium position at any time $\tau$ gives the electric field (Eq.({\ref{poissonL})) at the location of the sheet at that time.  This is true provided the ordering of sheets is maintained\cite{dawson_sheet}. 
Eqs.(\ref{mom_xL}) and (\ref{mom_yL}) are respectively the $x$ and $y$ component of equation of motion of the oscillating sheets. The above set of equations (Eqs.(\ref{continuity_L}) - (\ref{poissonL})) constitute the Dawson sheet model description of upper-hybrid oscillations. It is clear from above that with the knowledge of $\xi(x_0,\tau)$, evolution of large amplitude upper hybrid oscillations can be studied in terms of oscillating motion of electron sheets about their equilibrium positions. In the following sections, \ref{homogeneous} and \ref{inhomogeneous}, we respectively present exact solutions of the sheet equations with homogeneous and inhomogeneous magnetic field.
\begin{figure}
\centering
\includegraphics[width=1.0\linewidth,height=0.5\linewidth]{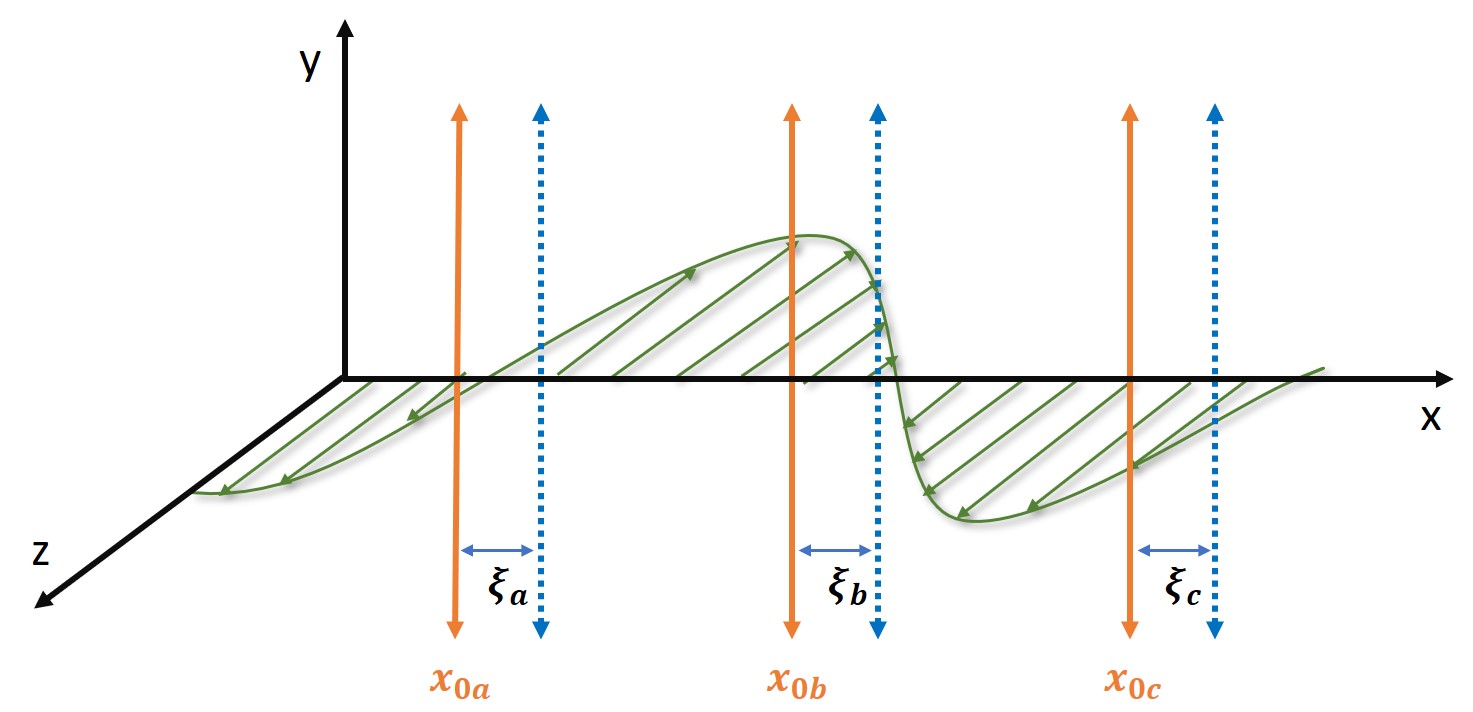}
\caption{Dawson sheet model; here orange solid lines represent the equilibrium postions of the electron sheets and the blue dotted lines represent the displaced postion of the electron sheets. The external magnetic field is directed along the z-axis. The green wavy line represent the inhomogeneity in the magnetic field along the x-axis, which is the case treated in section \ref{inhomogeneous}}
\label{sheet_model}
\end{figure}
\section{Homogeneous Magnetic Field} \label{homogeneous}
With an uniform external magnetic field, Eq. (\ref{mom_xL}) and (\ref{mom_yL}), may be combined to give a simple harmonic oscillator equation as follows

\begin{equation}\label{sheet_eqn}
\ddot{\xi} + \omega_{uh}^2 \xi = -\beta V_{y}(x_0)
\end{equation}
where $V_{y}(x_0) = v_{y0} - \beta \xi_0 $, $v_{y0} = v_y(x_0,0)$,   $\xi_0 = \xi(x_0,0)$ and $\omega_{uh}^2 = 1+\beta^2$.
Solving Eq.(\ref{sheet_eqn}), the expression for $\xi(x_0,\tau)$ may be written as 
\begin{widetext}
\begin{equation}\label{sheet_soln}
\xi(x_0,\tau) = \left[ \xi_0 + \frac{\beta}{\omega_{uh}^2}V_{y}(x_0) \right]   \cos\left(\omega_{uh} \tau\right)  + \frac{V_{x}(x_0)}{\omega_{uh}}\sin\left(\omega_{uh} \tau\right)  - \frac{\beta}{\omega_{uh}^2}V_{y}(x_0)
\end{equation}
where $V_{x}(x_0) = v_{x}(x_0,0)$.
Eq.(\ref{sheet_soln}) clearly shows that, even in the nonlinear case, electron sheets execute simple harmonic motion about their mean position with upper-hybrid frequency ($\omega_{uh}^2 = 1+\beta^2$). This, of course, is true, provided the ordering of electrons in the $x$ direction is maintained. Using the above expression for $\xi(x_0,\tau)$ and using Eqs.(\ref{continuity_L} - \ref{poissonL}), the expressions for electron density, $x$ and $y$ component of electron fluid velocity and electric field in Lagrange coordinates, may respectively be written as follows
%\begin{widetext}
\begin{eqnarray}
n_{e}(x_0,\tau) &=& \frac{1}{\left[ 
1 + \frac{\partial}{\partial x_0} \left\{ \left(\xi_0 + \frac{\beta}{\omega_{uh}^2} V_{y}(x_0) \right) \cos(\omega_{uh}\tau) +
\frac{V_{x}(x_0)}{\omega_{uh}}\sin(\omega_{uh} \tau) 
- \frac{\beta}{\omega_{uh}^2}V_{y}(x_0) \right\}
 \right]} \label{elec_den}\\
\nonumber \\
v_{x}(x_0,\tau) &=& -\omega_{uh}\left[ \xi_0 + \frac{\beta}{\omega_{uh}^2}V_{y}(x_0) \right]   \sin\left(\omega_{uh} \tau\right) + V_{x}(x_0) \cos(\omega_{uh} \tau) \label{velx}\\
\nonumber \\
v_{y}(x_0,\tau) &=& \beta \left[ \left\{\xi_0 + \frac{\beta}{\omega_{uh}^2}V_{y}(x_0) \right\}   \cos\left(\omega_{uh} \tau\right)  + \frac{V_{x}(x_0)}{\omega_{uh}}\sin\left(\omega_{uh} \tau\right) \right] + \frac{V_{y}(x_0)}{\omega_{uh}^2} \label{vely}\\
\nonumber \\
E(x_0,\tau) &=& \left[ \xi_0 + \frac{\beta}{\omega_{uh}^2}V_y(x_0) \right]   \cos\left(\omega_{uh} \tau\right)  + \frac{V_{x}(x_0)}{\omega_{uh}}\sin\left(\omega_{uh} \tau\right)  - \frac{\beta}{\omega_{uh}^2}V_{y}(x_0) \label{elec}
\end{eqnarray}
%\end{widetext}
and the transformation equations relating Lagrange to Euler coordinates may be written as 
%\begin{widetext}
\begin{eqnarray}
x &=& x_0 + \left[ \xi_0 + \frac{\beta}{\omega_{uh}^2}V_{y}(x_0) \right]   \cos\left(\omega_{uh} \tau\right)  + \frac{V_{x}(x_0)}{\omega_{uh}}\sin\left(\omega_{uh} \tau\right)  - \frac{\beta}{\omega_{uh}^2}V_{y}(x_0)\label{euler_lag_x}\\
\nonumber \\
t &=& \tau \label{euler_lag_t}
\end{eqnarray}
\end{widetext}
Therefore with the knowledge of $\xi_0 = \xi(x_0,0)$, $V_{x}(x_0)$ and $V_{y}(x_0)$, which may respectively be obtained from initial electron density and velocity profiles, Eqs.(\ref{elec_den} - \ref{elec}) along with the transformation relations Eq.(\ref{euler_lag_x}) and (\ref{euler_lag_t}) completely describe the spatio-temporal ($x,t$) evolution of large amplitude upper-hybrid oscillations in a cold plasma with spatially uniform external magnetic field. As mentioned above, this description of coherent upper hybrid oscillations holds provided the ordering of electron sheets is maintained at all times {\it i.e} 
$(\partial \xi / \partial x_0) > -1 $\cite{dawson59}.
This is directly related to the positivity of electron number density at all times (Eq.(6)). It thus follows from Eq.(\ref{sheet_soln}) that coherent oscillations can exist only if the initial conditions satisfy the following inequalities
\begin{widetext}
\begin{eqnarray}
n_e(x_0,0)\left[ 1 - \beta \frac{\partial v_y(x_0,0)}{\partial x_0}\right] &>& \frac{(1-\beta^2)}{2} \label{inequality1}\\
\nonumber \\
\left| n_e(x_0,0) \frac{\partial v_x(x_0,0)}{\partial x_0}\right| &<& \left[ 2\left\lbrace n_e(x_0,0) - \beta n_e(x_0,0) \frac{\partial v_y(x_0,0)}{\partial x_0} \right\rbrace - (1-\beta^2)  \right]^{1/2} \label{inequality2} 
\end{eqnarray}
\end{widetext}
The above inequalities are equivalent to the inequalities Eq.(52) and Eq.(53) presented in chapter 3 of reference\cite{Davidson72}; the equivalence may be shown in the same way as presented in reference\cite{rathee_2021} (in present article $\beta$, by definition, is a positive quantity whereas, in chapter $3$ of reference\cite{Davidson72} $\omega_{ce}$ is negative, by definition). Violation of the above inequalities lead to the breaking of upper hybrid oscillations. 
\section{Inhomogeneous Magnetic Field} \label{inhomogeneous}
In the presence of an inhomogeneous magnetic field of the form $B = B_0 \left[ 1 + \Delta \cos(\alpha x) \right]$, $\Delta$ being the amplitude of magnetic field inhomogeneity and $\alpha^{-1}$ being the magnetic field inhomogeneity scale length, the sheet equations (Eq.(\ref{mom_xL}) and Eq.(\ref{mom_yL})) respectively stand as
\begin{eqnarray}
\ddot{\xi} &=& -\xi - \beta\left[ 1+ \Delta\cos \alpha \left(x_0 +\xi\right) \right] v_y\\ \label{mom_xL_inh}
\nonumber \\
\dot{v}_y &=& \beta\left[ 1+ \Delta\cos \alpha \left(x_0 + \xi\right)  \right] \dot{\xi} \label{mom_yL_inh}
\end{eqnarray}
where $x = x_0 + \xi(x_0,\tau)$ has been used. Combining the above equations, we get 
\begin{widetext}
\begin{equation}\label{sheet_eqn_inh}
\ddot{\xi} + \omega_{uh}^2 \xi = - \beta^2\left[ \frac{V_{y}(x_0)}{\beta} + \Delta\left\{ \frac{V_{y}(x_0)}{\beta} \cos\alpha(x_0 + \xi) + \frac{1}{\alpha}\sin\alpha(x_0 + \xi) + \xi\cos\alpha(x_0 + \xi)\right\} +
\frac{\Delta^2}{2 \alpha} \sin 2\alpha(x_0 + \xi) \right] 
\end{equation}
%\end{widetext}
where $V_y(x_0) = v_{y0} - \beta\left[\xi_0 + (
\Delta/\alpha)\sin \alpha(x_0 + \xi_0) \right]$. 
For $\Delta = 0$, Eq.(\ref{sheet_eqn_inh}) reduces to Eq.(\ref{sheet_eqn}). 
%%
%where $\xi_0$ is given by the solution of the transcendental equation 
%$\xi_0 + \delta \sin k (x_0 + \xi_0) = 0$. Treating $(x_0,\xi)$ as parameters, the exact solution may be reduced to a quadrature as follows:
%\begin{eqnarray}
%x &=& x_0 + \xi \\ \label{x_old}
%\nonumber\\
%t &=& \int_{\xi_0}^{\xi} d\xi^{\prime} \Big[ g\left(x_0,\xi^{\prime} \right) \Big] ^{-\frac{1}{2}}\\ \label{t_old}
%\nonumber\\
%n_e &=& \left[ 1+\frac{\partial \xi}{\partial x_0}\right]^{-1} \label{density_old}
%\end{eqnarray}
%where
%\begin{equation}\label{deno_old}
%\frac{\partial \xi}{\partial x_0} =  \sqrt{\frac{g(x_0,\xi)}{g(x_0,\xi_0)}}\frac{\partial \xi_0}{\partial x_0} + \frac{1}{2}\sqrt{g(x_0,\xi)} \int^{\xi}_{\xi_0}  d \xi^{\prime} \frac{1}{[g(x_0,\xi^{\prime})]^{3/2}} \frac{\partial}{\partial x_0} g(x_0,\xi^{\prime})
%\end{equation}
%Eqs(\ref{x_old} - \ref{deno_old}) is the exact solution representing spatio-temporal evolution of electron plasma density in the inhomogeneous case, in parametric form $x=x(x_0,\xi)$, $t=t(x_0,\xi)$ and $n=n(x_0,\xi)$.
%
Using initial density as  $n_e(x,0) = n_0\left[1+\delta \cos(x) \right]$ 
%= n_0\left[ 1 + \delta \cos k_L(x_0 + \xi_0) \right]$ 
in Eq.(\ref{poissonL}) gives $\xi_0 = -\delta\sin(x_0 + \xi_0)$. Here $\delta$ is the amplitude of density perturbation and $x \rightarrow k_L x$, $k_L^{-1}$ being the scale length of density perturbation. The initial velocity profile is taken as $ v_x(x_0,0) = v_y(x_0,0) = 0$. 
%\begin{equation}\label{fi}
%\dot{\xi}^2=g(x_0,\xi) 
%\end{equation}
%where
%\begin{align}\label{fi_exp}
%g(x_0,\xi)  &=-\omega_{uh}^2 (\xi^2 - \xi_0^2) + 2\beta^2 \xi_0(\xi - \xi_0) - \frac{2\beta^2\Delta}{\alpha}(\xi - \xi_0)\left[ \sin \alpha(x_0 + \xi) - \sin \alpha(x_0  + \xi_0) \right]  \nonumber \\ 
%&\qquad{}  +\frac{\beta^2\Delta^2}{2\alpha^2}\left\lbrace \cos 2\alpha(x_0+\xi) + 2\cos \alpha(\xi - \xi_0) - 2 + \cos 2\alpha (x_0 + \xi_0) - 2\cos \alpha(2x_0 + \xi + \xi_0)\right\rbrace 
%\end{align}
%
Now redefining Lagrange coordinate as $x_l = x_0 + \xi_0$, and $\phi=\xi-\xi_0$, the first integral Eq. (\ref{sheet_eqn_inh}) gives 
%Eqs.(\ref{fi} - \ref{fi_exp}) may be rewritten as 
%
%\begin{align}\label{eqn10}
%\ddot{\phi} &= -(1+\beta^2) \phi - \beta^2\gamma\Big[\phi\cos(\alpha(x_l+\phi))+\frac{1}{\alpha}\left\lbrace \sin(\alpha(x_l+\phi))-\sin(\alpha x_l) \right\rbrace  \nonumber \\ 
%%&\qquad{} +\frac{\gamma}{2\alpha}  \left\lbrace \sin(2\alpha(x_l+\phi))+\sin(\alpha\phi)-\sin(\alpha(2x_l+\phi)) \right\rbrace  \Big]+\delta\sin(x_l)
%%\end{align}
%
\begin{equation}\label{phii}
\dot{\phi}^2=f\left(x_l,\phi \right) 
\end{equation}
where
\begin{align}\label{phii_exp}
f\left(x_l,\phi \right)  &=-(1 + \beta^2) \phi^2 + 2\phi\delta\sin(x_l) - \frac{2\beta^2\Delta\phi}{\alpha}(\sin(\alpha(x_l+\phi)-\sin(\alpha x_l))  \nonumber \\ 
&\qquad{}  +\frac{\beta^2\Delta^2}{2\alpha^2}\left\lbrace \cos(2\alpha(x_l+\phi))+2\cos(\alpha\phi)-2+\cos(2\alpha x_l)-2\cos(\alpha(2x_l+\phi)) \right\rbrace 
\end{align}
Finally treating these redefined variables $(x_l, \phi)$ as parameters, the expression for electron density ( Eq.(\ref{continuity_L}) ) can be reduced to a quadrature as follows, 
\begin{eqnarray}
x &=& x_l + \phi \label{x} \\
\nonumber\\
t &=& \int_{0}^{\phi} d\phi^{\prime} \Big[ f\left(x_l,\phi^{\prime} \right) \Big] ^{-\frac{1}{2}} \label{t} \\
\nonumber\\
n_e &=&\frac{1+\delta \cos(x_l)}{\left[ 1+\frac{\partial \phi}{\partial x_l}\right]} \label{density}
\end{eqnarray}
where
%
%\begin{equation}\label{deno}
%\frac{\partial \phi}{\partial x_l} =  [f(x_l,\phi)]^{1/2}
%\int^{\phi}_{0}  \frac{\delta \phi^{\prime} \cos(x_l) - \beta^2 \Delta \phi^{\prime} \left\{ (\cos\alpha(x_l + \phi^{\prime}) - cos(\alpha x_l) \right\} - \frac{\beta^2 \Delta^2}{2 \alpha^2}\left\{ \sin 2\alpha(x_l + \phi^{\prime}) + \sin(2 \alpha x_l) - 2 \sin\alpha(2 x_l + \phi^{\prime}) \right\}}{[f(x_l,\phi^{\prime}]^{3/2}} d\phi^{\prime} 
%\end{equation}
%
\def\x{\begin{multlined}[t][0.68\linewidth]
 \frac{1}{ \big[ f(x_l,\phi^{\prime})\big] ^{3/2}} \Bigg[\delta \phi^{\prime} \cos(x_l) - \beta^2 \Delta \phi^{\prime} \left\{ (\cos\alpha(x_l + \phi^{\prime}) - cos(\alpha x_l) \right\}         \\    
  - \frac{\beta^2 \Delta^2}{2 \alpha^2}\left\{ \sin 2\alpha(x_l + \phi^{\prime}) + \sin(2 \alpha x_l) - 2 \sin\alpha(2 x_l + \phi^{\prime}) \right\} \Bigg]  d\phi^{\prime}
    \end{multlined}}
   
\begin{equation}\label{deno}
    \frac{\partial \phi}{\partial x_l} =  \big[f(x_l,\phi)\big]^{1/2} \int^{\phi}_{0}\x
\end{equation}

\end{widetext}
Eq.(\ref{x} - \ref{deno}) represent spatio-temporal evolution of electron density in the inhomogeneous case, in parametric form $x=x(x_l,\phi)$, $t=t(x_l,\phi)$ and $n=n(x_l,\phi)$.  The nature of the solution may be seen from fig.(\ref{phase_space}) where the phase space trajectories $(\phi, \dot{\phi})$ have been plotted for different values of $x_l$ ( initial positions of different electron sheets). It is clear from fig.(\ref{phase_space}), that although motion of each electron sheet is periodic in nature, its time period is dependent on its initial spatial position\cite{infeld89,chandanpre}. 
\begin{figure}
\includegraphics[width=1.0\linewidth,height=0.5\linewidth]{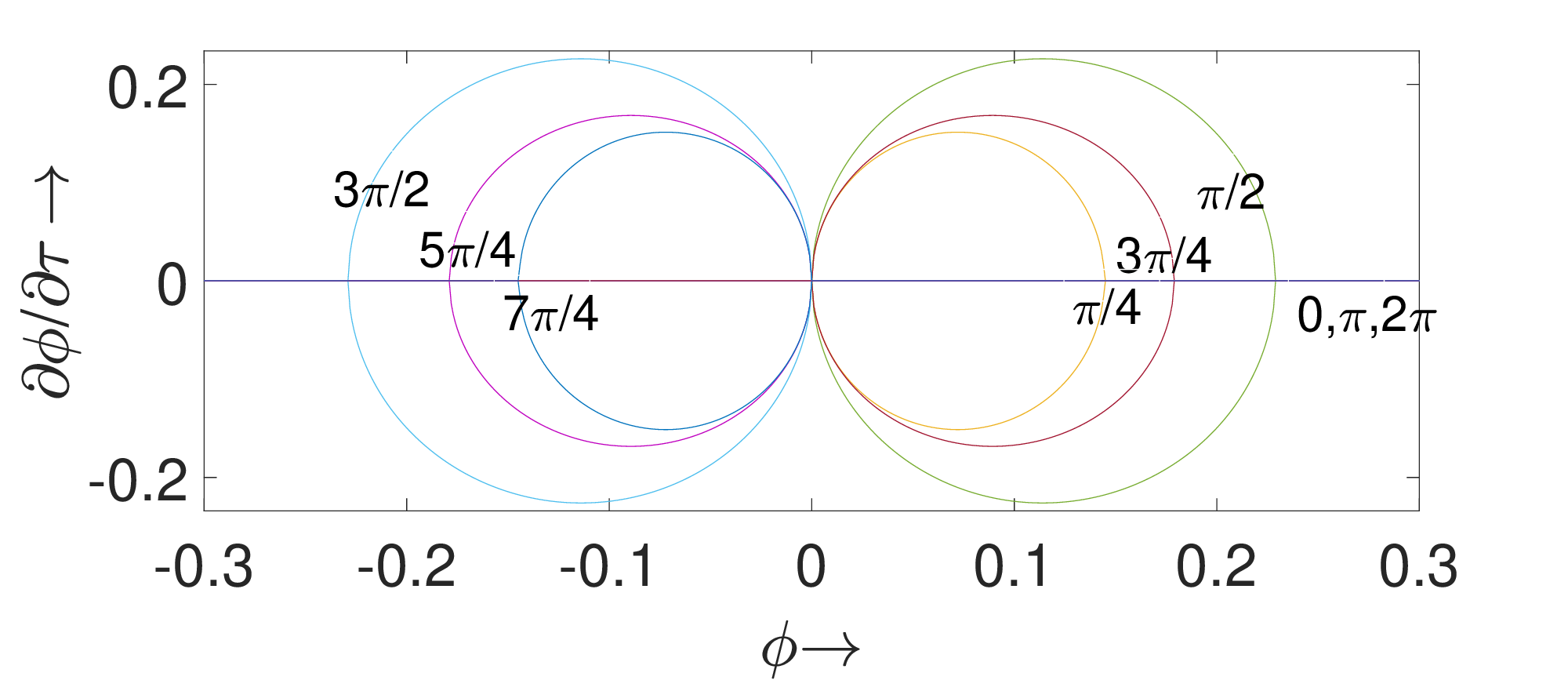}
\caption{Phase space plot ($\phi,\dot{\phi}$) for $\delta = 0.45$, $\Delta = 0.1$, $\alpha = 1.0$, and $\beta^2 = 3$}
\label{phase_space}
\end{figure}
%
%{\bf This time period may be represented as an integral of Eq.(\ref{t}) between the two turning points ($\dot{\phi}=0$), from which it is evident that the frequency of oscillation is in general a function of $x_l$ ({\it i.e.} initial position of electron sheets).}
% 
In order to physically illustrate the consequence of this, in the following subsection we present approximate expressions for frequency of oscillation of an electron sheet and electron number density explicitly showing their dependence on ($x_l$, $\tau$).
\subsection{Approximate Analysis}\label{app_analysis}
In order to make the solution physically transparent, we express the spatio-temporal evolution of electron density approximately in terms of ($x_l$, $\tau$). For this purpose, we obtain an approximate solution for $\phi$ by linearising Eq.(\ref{sheet_eqn_inh}), as follows
\begin{align}
\ddot{\phi}+\omega^2\phi &\approx \delta\sin(x_l)\label{linear_phi}
\end{align}
which gives
\begin{equation}\label{linear_soln}
\phi(x_l,\tau) \approx \frac{\delta\sin(x_l)}{\omega^2}\Big[ 1-\cos(\omega\tau)\Big] 
\end{equation}
where $\omega$ is given by 
\begin{equation}\label{omega}
\omega^2 = 1+\beta^2 \left[ 1 + \Delta \cos(\alpha x_l)\right]^2 
\end{equation}
Using Eq.(\ref{density}) and Eq.(\ref{linear_soln}), electron density may be expressed in terms of ($x_l$, $\tau$) as
\begin{widetext}
\begin{equation}\label{approx_den}
n_e(x_l,\tau) \approx \left( 1+\delta\cos( x_l) \right) \Bigg[1+\left[ \frac{\delta}{\omega^2}\cos(x_l) -\frac{2\delta}{\omega^3}\frac{d\omega}{d x_l}\sin(x_l)\right] \left\lbrace  1-\cos(\omega\tau) \right\rbrace +\frac{\tau \delta}{\omega^2}\frac{d\omega}{d x_l} \sin(x_l)\sin(\omega\tau)  \Bigg]^{-1}  
\end{equation}
\end{widetext} 
The physics contained in the above equations may be explained as follows.
From the expression of $\phi(x_l,\tau)$, we see that electron sheets oscillate with frequency $\omega$, 
which itself is a periodic function of $x_l$. This is clearly seen in Fig.\ref{fig1}, where we show the comparison of the approximate frequency 
Eq.(\ref{omega}) with the exact frequency of oscillation obtained by numerically integrating Eq.(\ref{t}) between two turning points {\it i.e.} ($\dot{\phi}=0$). 
\begin{figure}
\includegraphics[width=1.0\linewidth,height=0.5\linewidth]{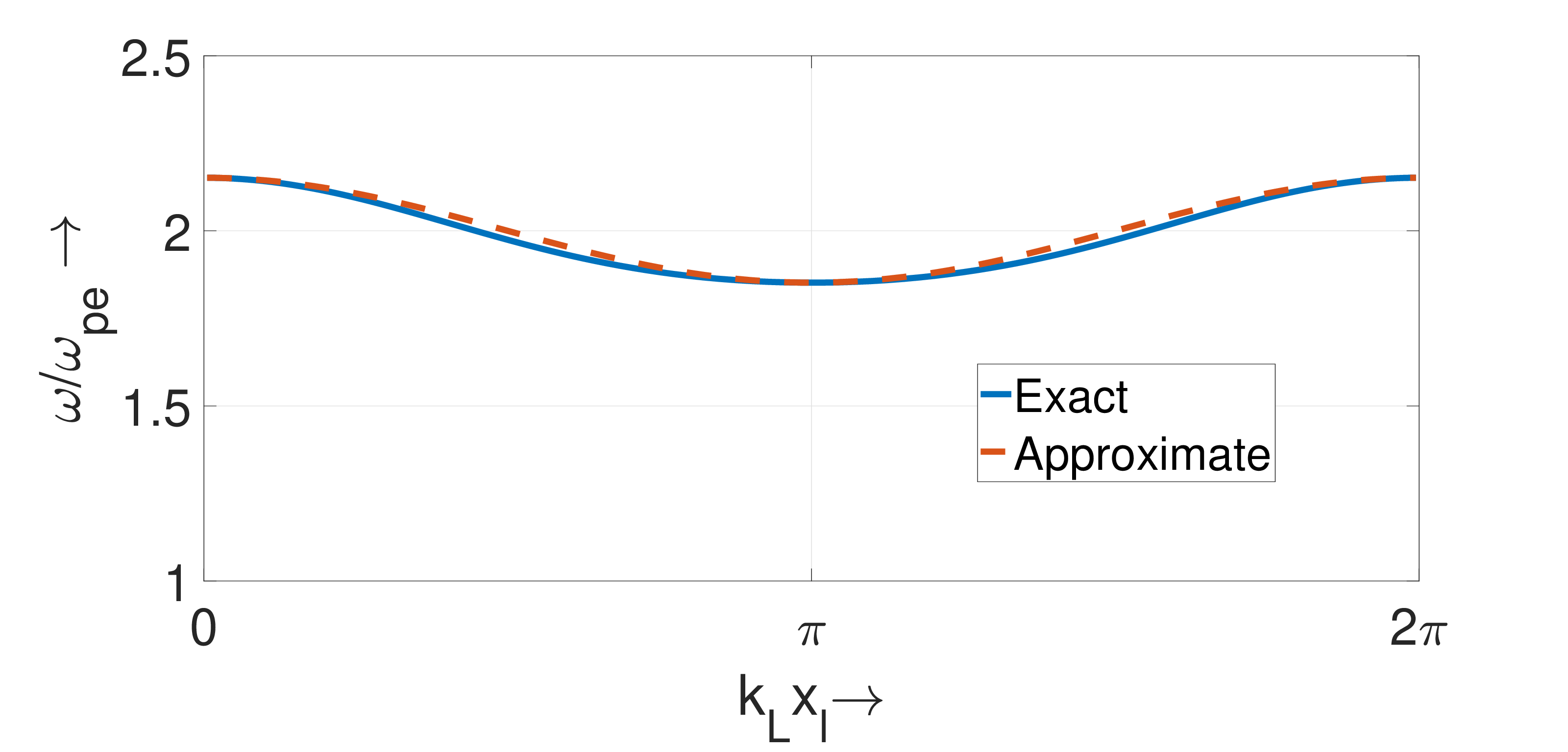}
\caption{Exact(blue) and Approximate(red) frequency as a function of initial position for $\delta = 0.45$, $\Delta = 0.1$, $\alpha=1.0$ and $\beta^2 =3$}
\label{fig1}
\end{figure}
Since any coherent mode is made up of a large number of electron sheets oscillating about their equilibrium positions, this spatial dependency of $\omega$ causes the neighbouring electron sheets to gradually go out of phase with time. This eventually leads to crossing of electron sheet trajectories resulting in singularities in the electron density profile. This is the phenomenon of phase mixing leading to wave breaking. The time at which the density becomes singular may be estimated from the Eq.(\ref{approx_den}), whose denominator vanishes approximately when $(\tau \delta/ \omega^2) d \omega / d x_l \rightarrow 1$. This gives the phase mixing (or wave breaking) time as 
\begin{equation}\label{phase_mixing}
\omega_{pe}\tau_{mix} \approx \frac{\left( 1+\beta^2 \right)^{3/2} \, k_L}{\beta^2 \, \delta \, \Delta \, \alpha }
\end{equation}
which shows that for given magnetic field strength ($\beta$), the phase-mixing time is inversely dependent on density perturbation amplitude ($\delta$), amplitude of magnetic field inhomogeneity ($\Delta$) and directly on the ratio of magnetic field inhomogeneity scale length to imposed density perturbation scale length ($(\alpha/k_L)^{-1}$). 

\section{Numerical Results} \label{numerical}
\begin{figure}
\includegraphics[width=1.0\linewidth,height=0.5\linewidth]{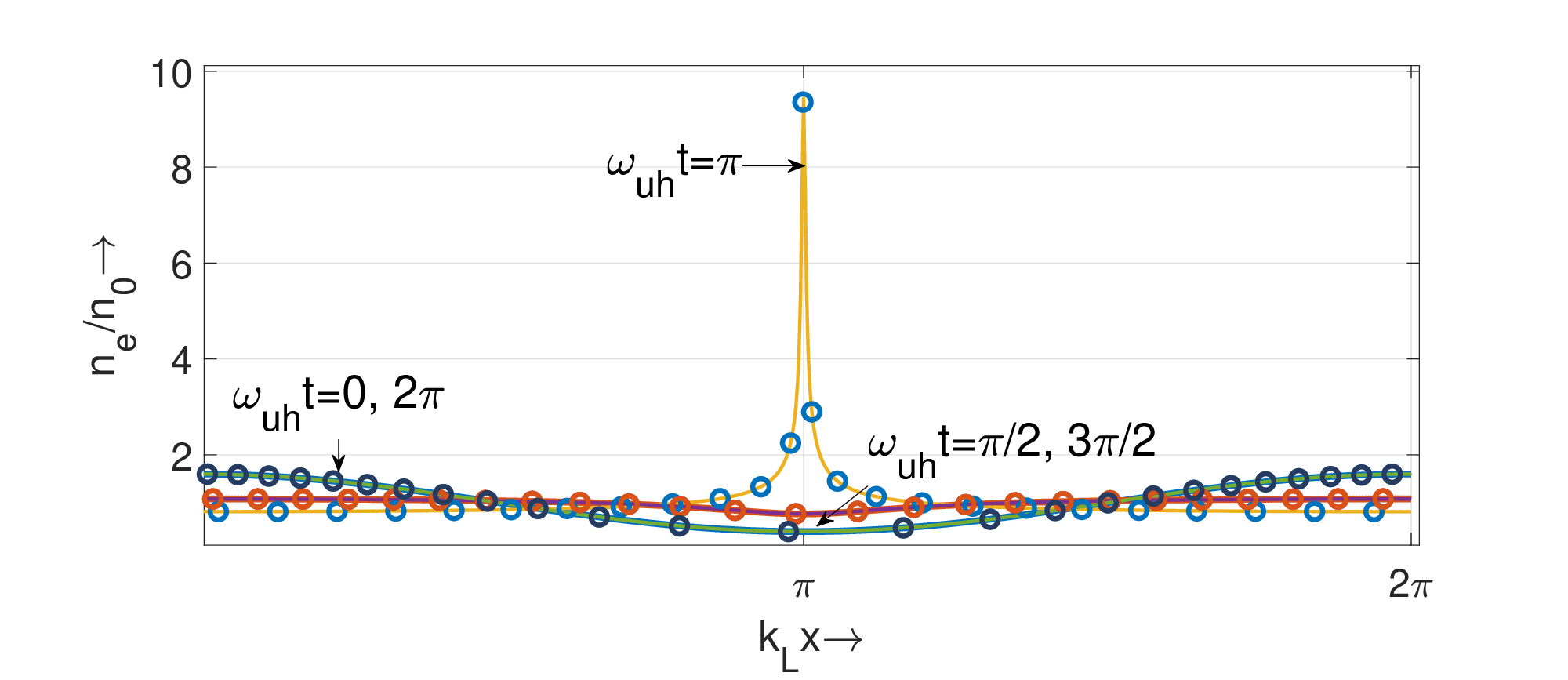}
\caption{$n_e/n_0$ as a function of space at different times for $\delta = 0.6$ and $\beta = 0.5$, where $\omega_{uh} = \sqrt{1+\beta^2}$. Here solid lines represent simulation results and dots represent theoretical values.}
\label{fig:1}
\end{figure}
In this section, we numerically demonstrate the spatio-temporal evolution of upper-hybrid oscillations using a sheet code based on Dawson sheet model. We present results corresponding to two sets of simulations: the first set deals with the standard case of upper-hybrid oscillations in a homogeneous external magnetic field, and in the second set of simulations, we study the affect of inhomogeneity in the external magnetic field on upper-hybrid oscillations.   

In our simulations, we have followed the motion of $\sim 4 \times 10^4$ electron sheets. For given initial electron density and velocity profiles, and using periodic boundary conditions, the electron sheets are initially placed in phase space following the technique of inversion of cumulative distribution function \cite{birdsall85}. The equation of motion for each electron sheet is then solved using Boris algorithm \cite{birdsall85} and the electron sheets are tracked for several hundreds of plasma periods.
At each time step, ordering of sheets is checked for sheet crossing. In the homogeneous case, if the initial electron density and velocity profiles satisfy the inequalities given by (Eqs. (\ref{inequality1}) and (\ref{inequality2})), the sheets, as expected, do not cross during the evolution of the upper-hybrid mode, whereas for the case with inhomogeneous magnetic field, as discussed above,  where the upper-hybrid frequency acquires spatial dependence, crossing of electron sheets eventually occurs during the evolution of the upper-hybrid mode. This results in breaking of the upper-hybrid oscillation in the inhomogeneous magnetic field case; and the time at which neighbouring electron sheets cross is taken as the phase mixing ( wave breaking ) time.  We terminate our code at this time because the expression for self-consistent electric field ($E = 4\pi en_0 \xi$) used in equation of motion (Eq.(\ref{mom_xL})) becomes invalid beyond this point. Finally, to study the spatio-temporal evolution of electron density profile, we superimpose a spatial grid on the oscillating electron sheets, thereby dividing the whole simulation domain into  cells. At each time step, the electron number density is recorded at the cell centres using cloud-in-cell method\cite{birdsall85}.

For the first set of simulations, we excite upper hybrid oscillations by perturbing the electron density with a sinusoidal perturbation, $n_e(x,0) = \left[1+\delta \cos(x) \right]$, with amplitude $\delta = 0.6$ and  $ v_x(x,0) = v_y(x,0) = 0$, {\it i.e.} without any initial velocity shear. The ratio of electron cyclotron frequency to plasma frequency $\beta$ is chosen as $\beta = 0.5$. These parameters are chosen in a way, such that the inequalities Eq. (\ref{inequality1}) and Eq. (\ref{inequality2}) are satisfied ( $\mid \delta n / n_0 \mid < (1 + \beta^2)/2$ ). Fig.\ref{fig:1} shows the spatio-temporal evolution of electron density profile. Here the solid lines represent simulation results and dots represent theoretical profile obtained using Eqs.(\ref{elec_den}), Eq.(\ref{euler_lag_x}) and Eq.(\ref{euler_lag_t}). The figure shows a good match between theory and simulation, thus clearly validating our numerical code. It may be seen from this figure that the electron density returns to the original profile after one oscillation and the frequency of this oscillation is $\omega_{uh}$.
%
%\begin{widetext}
\begin{figure}
     \centering
     \begin{subfigure}[b]{0.48\textwidth}
         \centering
         \includegraphics[width=1.0\linewidth,height=0.5\linewidth]{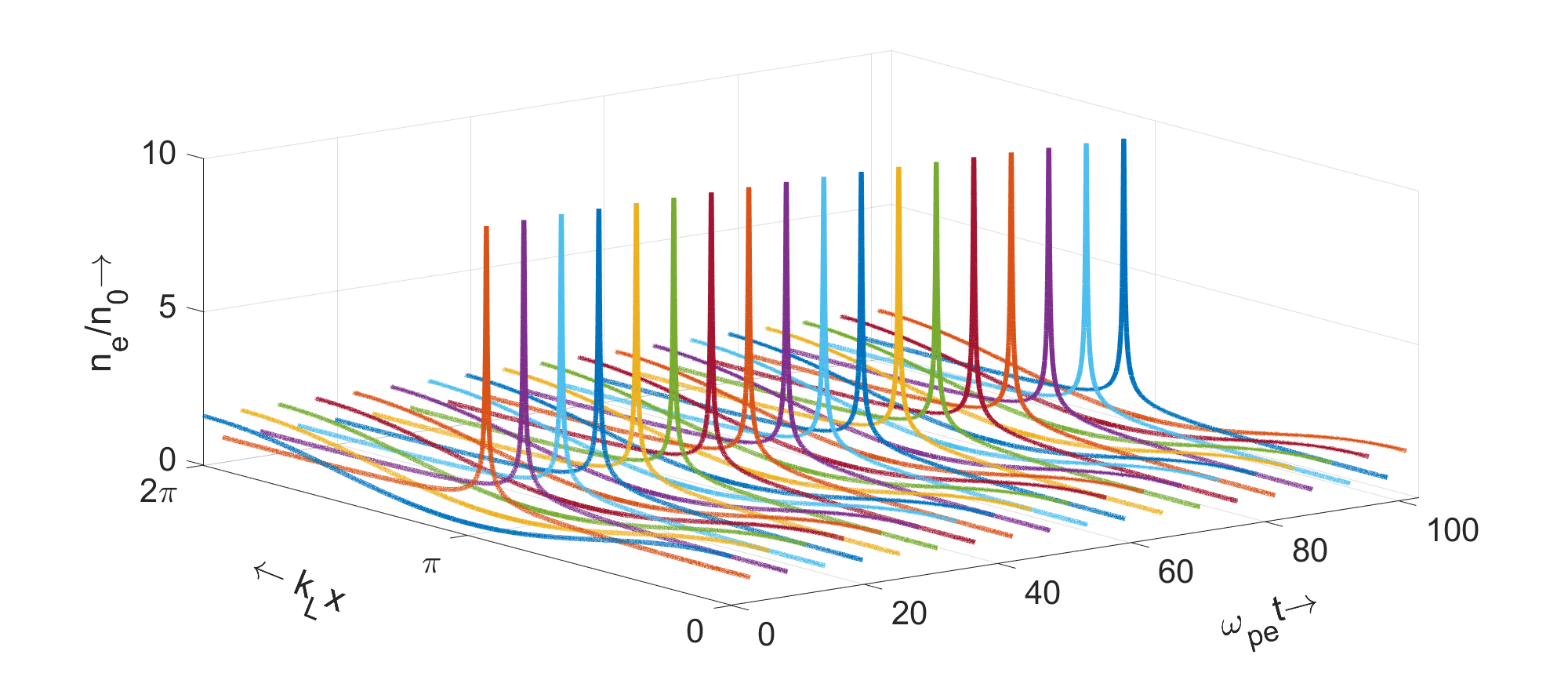}
\caption{For homogeneous external magnetic field with $\delta=0.6$ and $\beta = 0.5$}
\label{fig3anew}
     \end{subfigure}
%     \hfill
     \begin{subfigure}[b]{0.48\textwidth}
         \centering
\includegraphics[width=1.0\linewidth,height=0.5\linewidth]{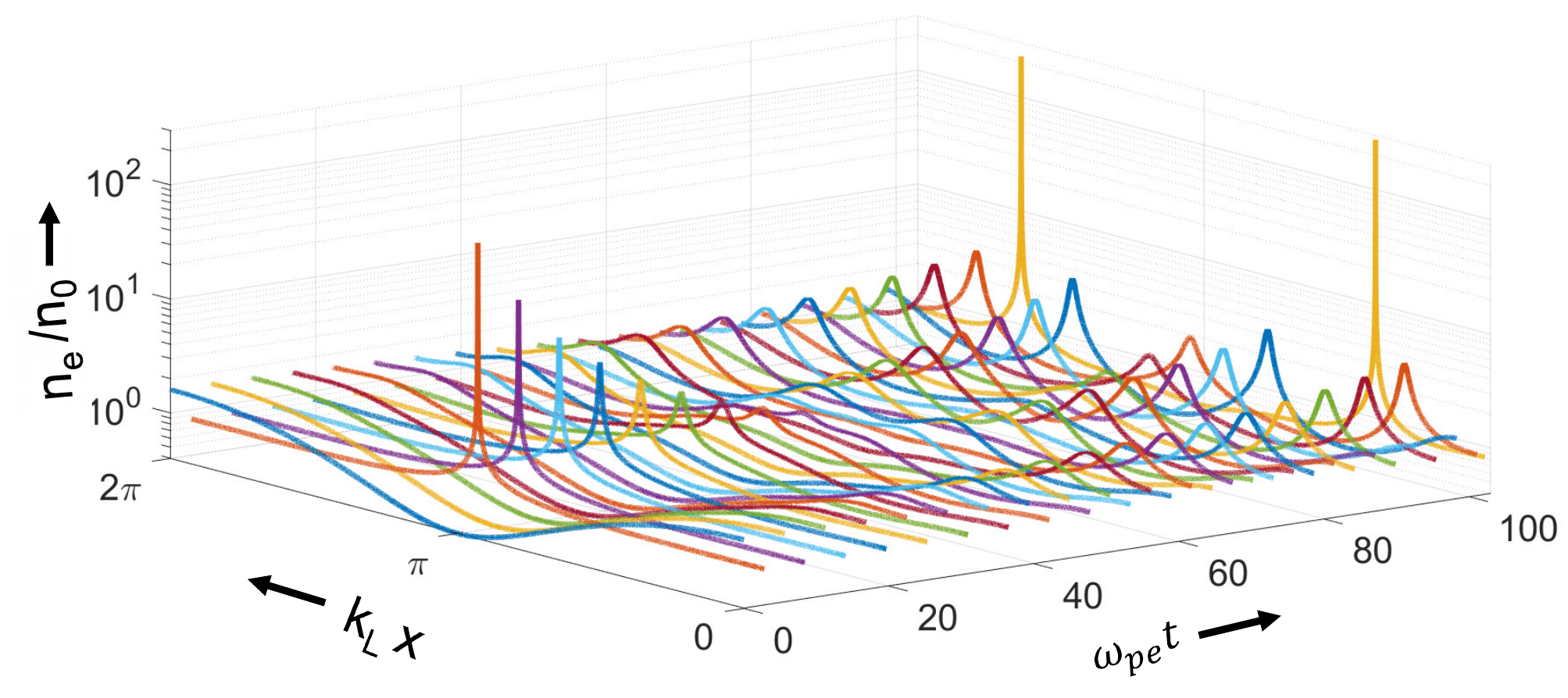}
\caption{For inhomogeneous external magnetic field with $\delta = 0.6$, 
        $\Delta = 0.1$, $\alpha = 1.0$ and $\beta = 0.5$}
\label{fig3bnew}
     \end{subfigure}
        \caption{Space-time evolution of density with homogeneous (\ref{fig3anew}) and inhomogeneous (\ref{fig3bnew}) external magnetic field}
        \label{fig3}
\end{figure}
\begin{figure}
     \centering
     \begin{subfigure}[b]{0.48\textwidth}
         \centering
%
%\includegraphics[width=1.0\linewidth,height=0.5\linewidth]{figure6a_uk.eps}
%\caption{ For homogeneous magnetic field with $\delta = 0.45$ and 
%$\beta^2 =3.0$ at $\omega_{p}t = 0.0$ (inset)  and $\omega_{p}t =  62.8319$.}
%
\includegraphics[width=1.0\linewidth,height=0.5\linewidth]{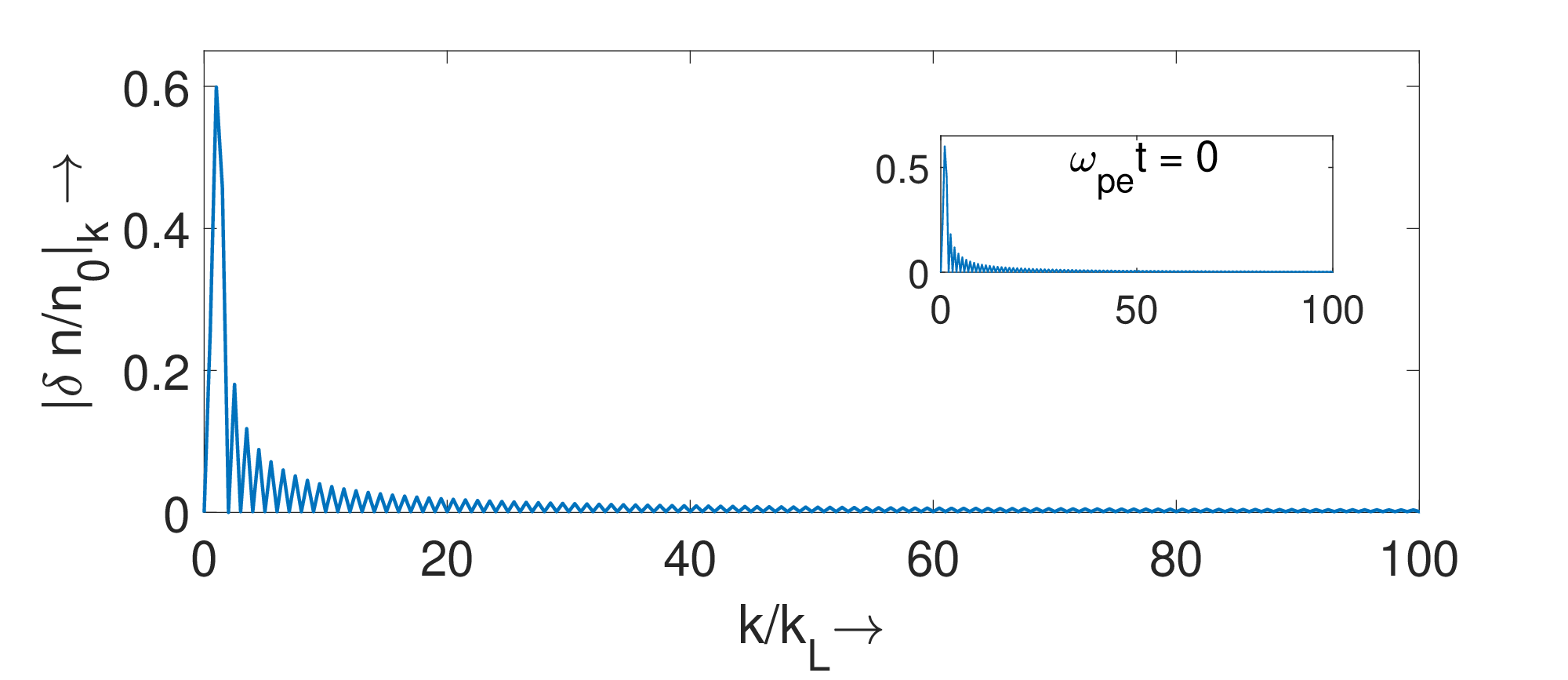}
\caption{ For homogeneous magnetic field with $\delta = 0.6$ and $\beta = 0.5$ at $\omega_{p}t = 0.0$ (inset)  and $\omega_{p}t =  95.5$}
\label{fig:3a}
     \end{subfigure}
     \hfill
     \begin{subfigure}[b]{0.48\textwidth}
         \centering
%
%\includegraphics[width=1.0\linewidth,height=0.5\linewidth]{figure6b_uk.eps}
%\caption{For inhomogeneous magnetic field with $\delta = 0.45$, $\Delta =0.1$, $\alpha=1.0$ and $\beta^2=3$ at $\omega_{p}t = 0.0$ (inset) and $\omega_{p}t = 60.2190$.}
%
\includegraphics[width=1.0\linewidth,height=0.5\linewidth]{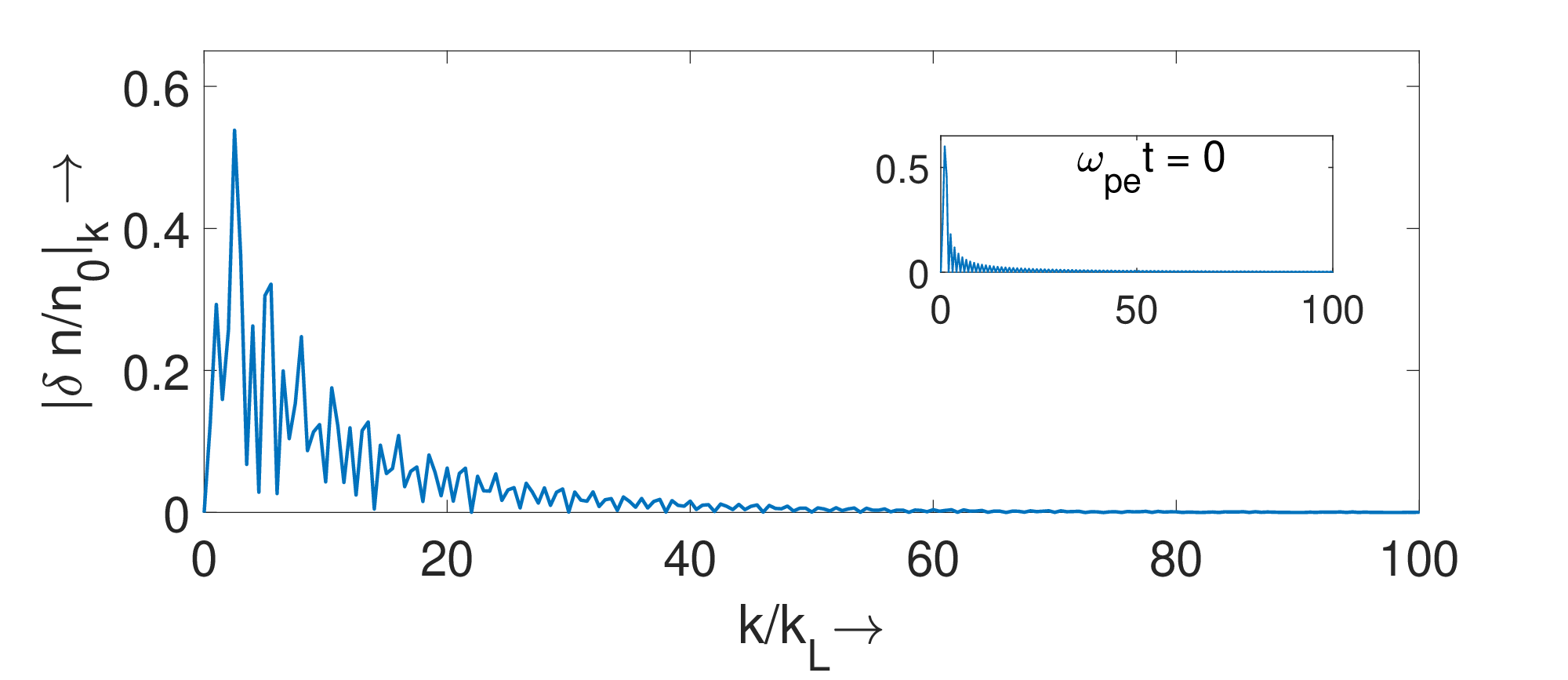}
\caption{For inhomogeneous magnetic field with $\delta = 0.6$, $\Delta = 0.1$, $\alpha=1.0$ and $\beta = 0.5$ at $\omega_{p}t = 0.0$ (inset) and $\omega_{p}t = 95.5$.}
\label{fig:3b}
     \end{subfigure}
        \caption{Fourier spectrum of Upper Hybrid Oscillations with homogeneous (\ref{fig:3a}) and inhomogeneous (\ref{fig:3b}) external magnetic field}
        \label{fig:3}
\end{figure}
\begin{figure}
%\includegraphics[width=1.0\linewidth,height=0.5\linewidth]{density_amplitude_corresponding_to_different_k.eps}
%\caption{Time evolution of first four density modes for homogeneous 
%(orange) with $\delta =0.45$, $\beta^2=3$ and inhomogeneous (blue) external magnetic field with $\delta=0.45$, $\Delta = 0.1$, $\alpha = 1.0$, $\beta^2=3$}
\includegraphics[width=1.0\linewidth,height=0.5\linewidth]{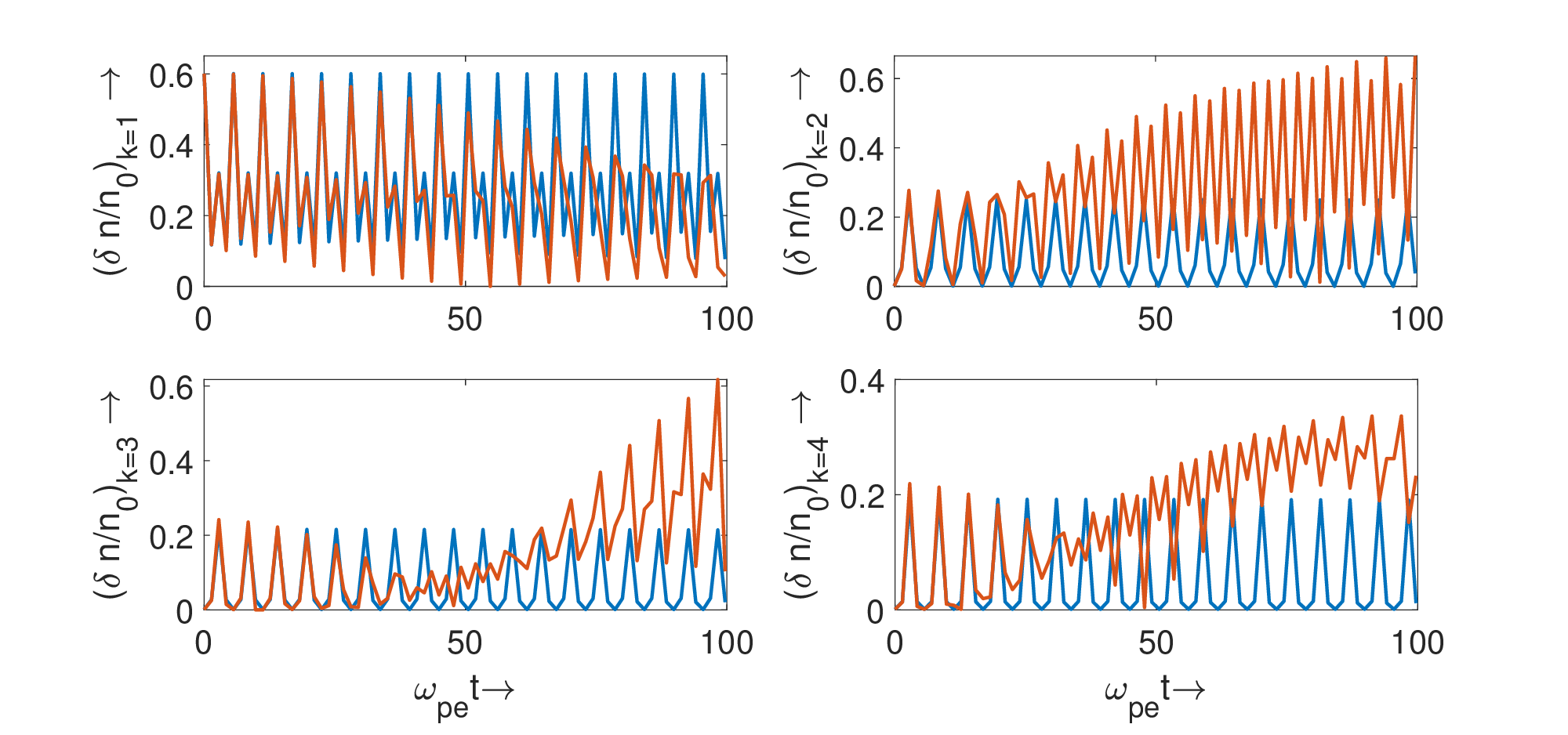}
\caption{Time evolution of first four density modes for homogeneous 
(blue) with $\delta =0.6$, $\beta=0.5$ and inhomogeneous (orange) external magnetic field with $\delta=0.6$, $\Delta = 0.1$, $\alpha = 1.0$, $\beta=0.5$}
\label{fig:4}
\end{figure}
%\end{widetext}

For the second set of simulations, we introduce a sinusoidal inhomogeneity in the external magnetic field of the form $B(x) = B_0 [1 + \Delta \cos(\alpha x) ]$ and study the affect of inhomogeneity on spatio-temporal evolution of upper-hybrid mode. This is in addition to the electron density and velocity perturbations introduced above. 
In Fig.(\ref{fig3}), we compare the spatio-temporal evolution of the upper-hybrid mode ( electron density profile ) in a homogeneous (Fig.\ref{fig3anew}) and in an inhomogeneous (Fig.\ref{fig3bnew}) external magnetic field. It is seen from Fig.(\ref{fig3anew}) that the electron density profile remains unaltered for hundreds of plasma periods; the energy which is initially loaded on a long wavelength mode, in each cycle, goes into higher harmonics and returns to the original mode. This continuous sloshing of energy back-and-forth between the original mode and higher harmonics without breaking happens in the homogeneous case only, whereas in fig.(\ref{fig3bnew}) as time progresses, the density profile becomes more and more spiky; the energy, which is initially loaded in a single mode, is irreversibly transferred to higher and higher harmonics. This is a signature of phase mixing. The inhomogeneity in external magnetic field thus destroys the coherent motion, and eventually leads to breaking of the upper-hybrid wave via phase-mixing. 

To further illustrate the phenomenon of phase-mixing, in Fig.(\ref{fig:3}), we present the Fourier spectrum of the electron density profile after several tens of plasma periods (the initial spectrum is presented in inset) for both homogeneous (Fig.\ref{fig:3a}) and inhomogeneous (Fig.\ref{fig:3b}) external magnetic field. The parameters chosen are the same as used for Fig.(\ref{fig3}). Fig.(\ref{fig:3a}) shows that the energy which is initially loaded on the primary mode remains in that mode even after several tens of plasma periods whereas inclusion of inhomogeneity with amplitude $\Delta = 0.1$ and with mode number 
$\alpha=1.0$, results in leakage of energy to higher Fourier modes ( see Fig.\ref{fig:3b} ), within the same time period. 
Fig.(\ref{fig:4}) shows the temporal evolution of the first four density modes for homogeneous (orange) and inhomogeneous (blue) magnetic field cases, for the same set of parameters as used in Fig.(\ref{fig:3}). It clearly shows the decay of the primary mode and the growth of higher harmonics with time when the external magnetic field is inhomogeneous.
\begin{figure}
\includegraphics[width=1.0\linewidth,height=0.5\linewidth]{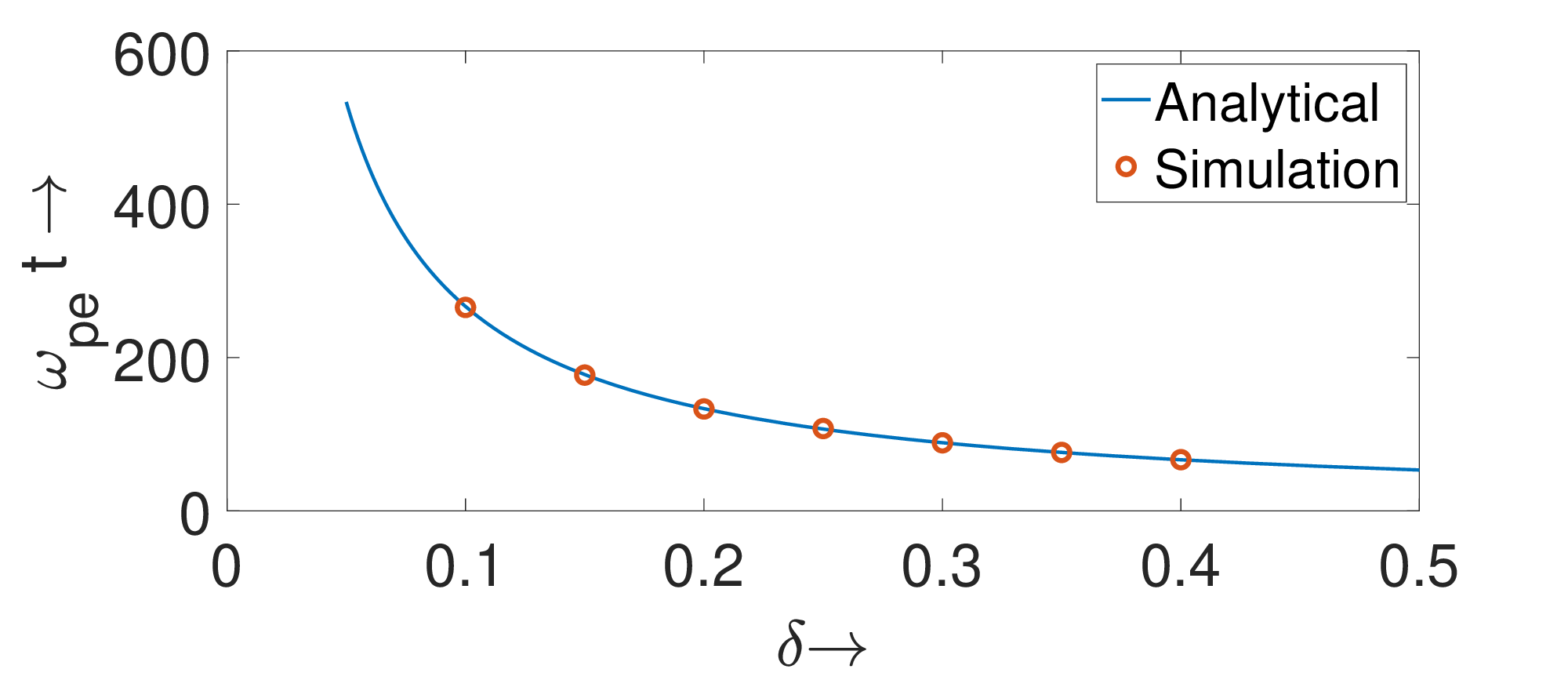}
\caption{$\tau_{mix}$ as a function of $\delta$, with $\Delta = 0.1$, 
$\alpha = 1.0$ and $\beta^2 = 3$ }
\label{fig7}
\end{figure}
 \begin{figure}
\includegraphics[width=1.0\linewidth,height=0.5\linewidth]{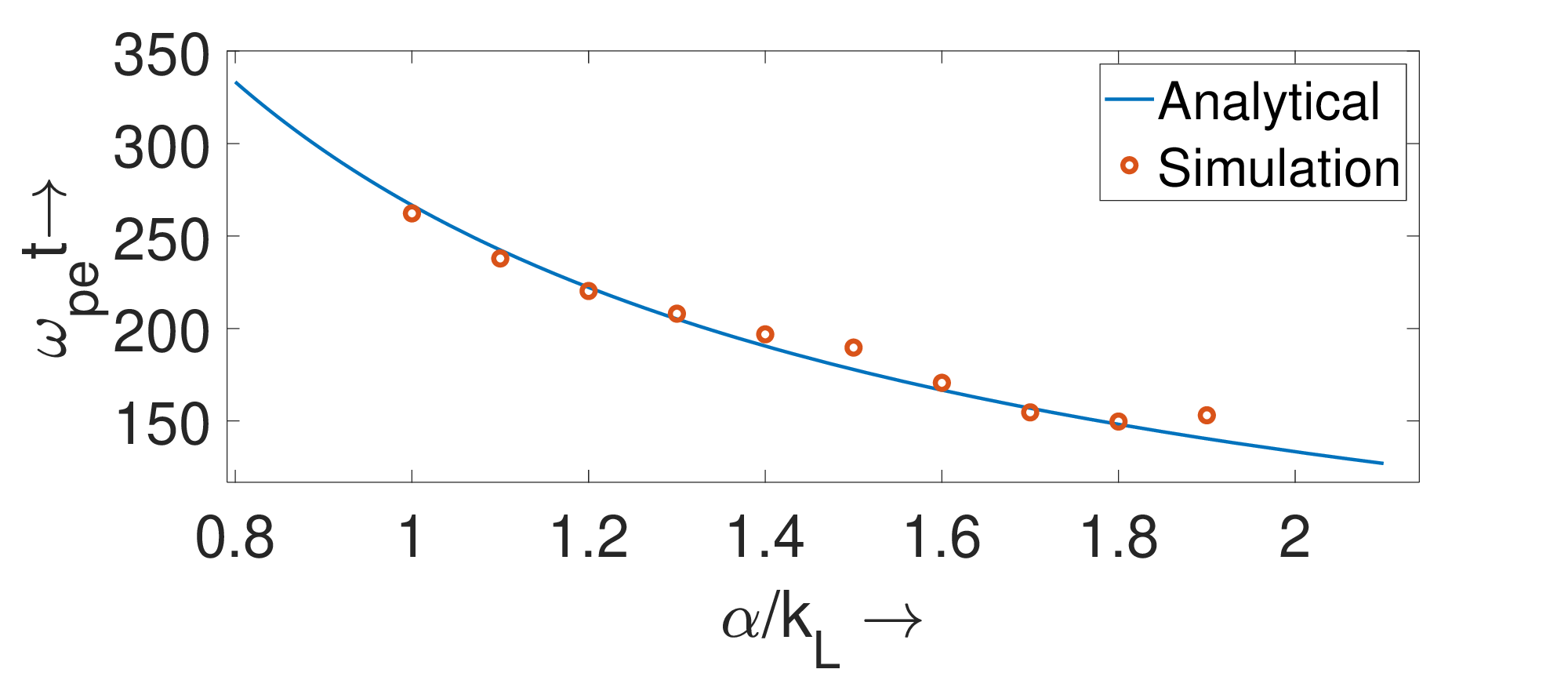}
\caption{$\tau_{mix}$ as a function of $\alpha$ with $\delta=0.1$, $\Delta = 0.1$, and $\beta^2 = 3$}
\label{fig8}
\end{figure}
\begin{figure}
\includegraphics[width=1.0\linewidth,height=0.5\linewidth]{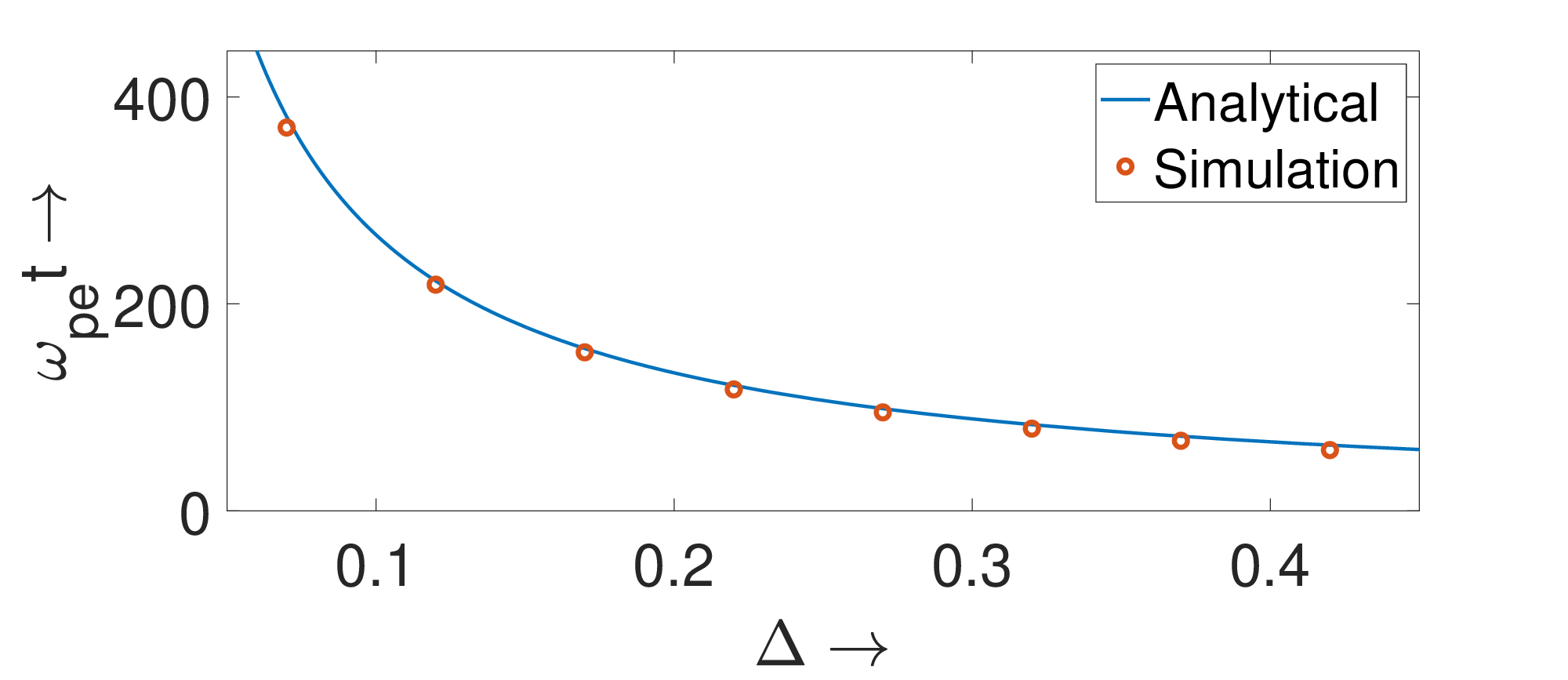}
\caption{$\tau_{mix}$ as a function of $\Delta$ with $\delta = 0.1$,
$\alpha = 1.0$ and $\beta^2 = 3$}
\label{fig9}
\end{figure}

Next we present the dependence of the wave-breaking/phase-mixing time on the density perturbation amplitude $\delta$, the magnetic field inhomogeneity scale length $\alpha$ and the magnetic field inhomogeneity amplitude $\Delta$ by varying one of the parameters, keeping the other two constant. Fig.(\ref{fig7}) - (\ref{fig9}) respectively show the dependence of phase mixing time on the density perturbation amplitude $\delta$ (keeping $\Delta = 0.1$ and $\alpha=1.0$ fixed), inverse of magnetic field inhomogeneity scale length $\alpha$ (keeping $\delta = 0.1$ and $\Delta = 0.1$ fixed) and magnetic field inhomogeneity amplitude $\Delta$ (keeping $\delta = 0.1$ and $\alpha = 1.0$ fixed). The dots represent the phase mixing time obtained from simulation by recording the sheet crossing time  and the solid line represent the theoretical scaling, given by 
Eq.(\ref{phase_mixing}). The simulation results which clearly support the phase mixing formula derived in section \ref{app_analysis}, shows that the phase mixing time scales inversely with the density perturbation amplitude $\delta$, and the magnetic field inhomogeneity amplitude 
$\Delta$ and directly with the magnetic field inhomogeneity scale length $\alpha^{-1}$. 

\begin{figure}
\includegraphics[width=1.0\linewidth,height=0.5\linewidth]{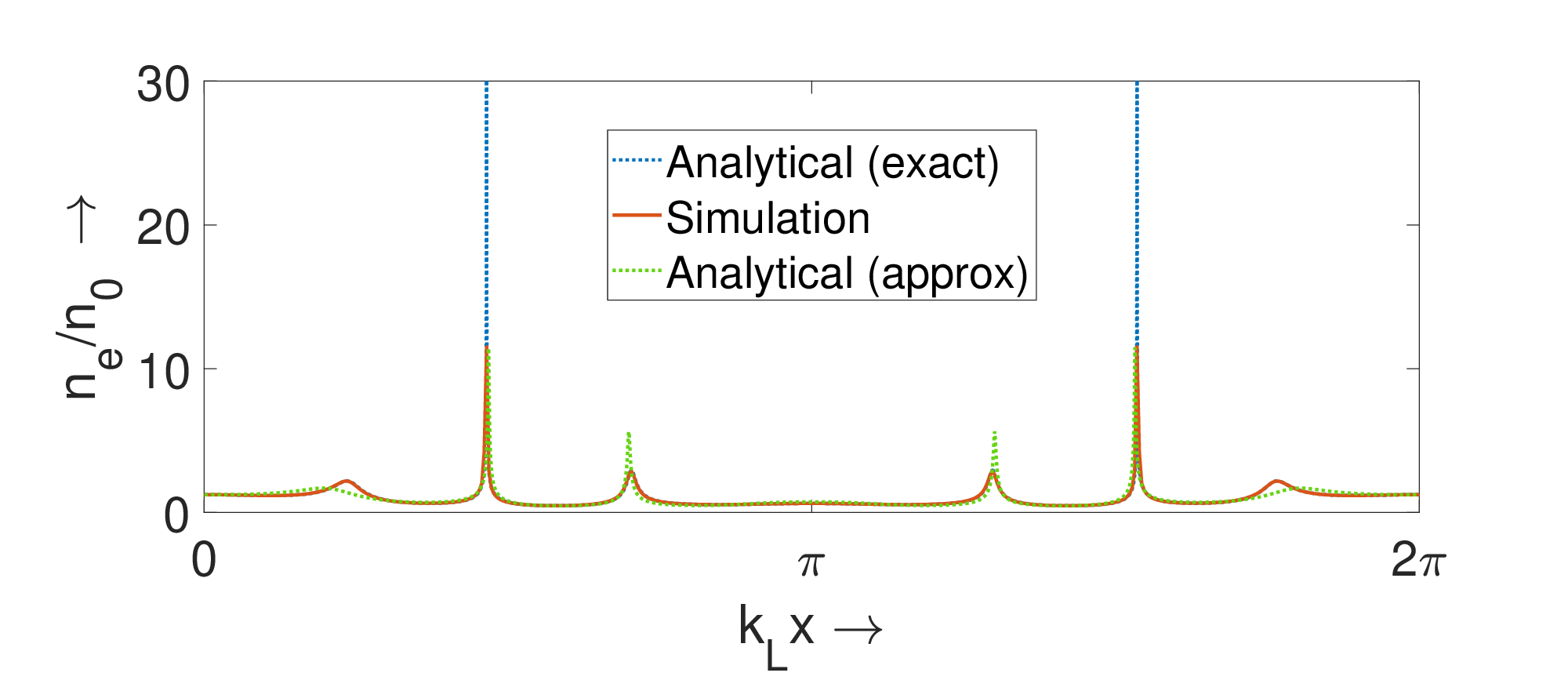}
\caption{ Normalized density as a function of position ( exact(blue), approximate (green), and simulation(orange) ) close to the phase mixing time $\omega_{pe}\tau \approx 59.3$, for $\delta = 0.45$, $\Delta=0.1$, $\alpha=1.0$ and $\beta^2 = 3$ }
\label{fig5}
\end{figure}
Finally in Fig.(\ref{fig5}), we present the comparison of analytically estimated ( both exact Eq.(\ref{density}) and approximate Eq.(\ref{approx_den})) electron density with the density obtained from simulation, at wave-breaking time (the time at which two adjacent electron sheet cross over). The method used to obtain the exact theoretical curve (blue) in Fig.(\ref{fig5}) is as follows. From Eq.(\ref{t}), at a given time $\tau$, we get values of $\phi$ corresponding to all values of $x_l$. Using these values of $(x_l,\phi)$, from Eq.(\ref{x}), we get the corresponding values of $x$. These values of $(x_l, \phi)$, when used in Eq.(\ref{density}), yields the corresponding values of density, which is then plotted as a function of $x$. This figure also shows that our simulations are in excellent agreement with the analytical results.
%
%Similarly, we get values of all plasma parameters corresponding to that set of $(x_l,\phi)$, using Eq.\ref{eqn15}. Hence we get plasma parameters as function of space at a given time.
%
\section{Summary}\label{sec4}
In conclusion, we have shown numerically using a 1-1/2 D sheet code that inclusion of homogeneous external magnetic field increases the critical amplitude of electron oscillations (upper-hybrid oscillations), \textit{i.e.} the initial amplitude of perturbation below which a cold plasma sustains coherent motion, increases with the increase in external magnetic field. It is further shown that in the presence of an inhomogeneous external magnetic field, upper hybrid oscillations break via the process of phase mixing at arbitrarily low amplitudes, and the breaking time inversely depends on the amplitude of density and magnetic field inhomogeneity and directly on the ratio of magnetic field inhomogeneity scale length to density perturbation scale length. As stated in the introduction, the study of electrostatic waves propagating transverse to an inhomogeneous magnetic field is a problem of interest from the viewpoint of particle acceleration experiments. It is well known that resonant particles trapped  in an electrostatic wave propagating transverse to a homogeneous magnetic field can in principle gain unbounded amount of energy.  This is the surfatron mechanism of acceleration\cite{katsouleas83prl}. Recently it has been shown that in the presence of an inhomogeneous magnetic field, even non-resonant particles can get accelerated\cite{artemyev2015prl}. This of course can only occur provided the electrostatic wave itself survives because of phase mixing effects, thus indicating the relevance of the present work. We have extended the present work on phase mixing of electrostatic waves in inhomogeneous magnetic field to include density inhomogeneities, finite electron temperature and relativistic effects. These will be presented in future publications.   

%The study of phase mixing of oscillations/waves in the presence of an inhomogeneous magnetic field is a problem of interest as inhomogeneous magnetic field is found in all realistic situations starting from laboratory plasma, tokamak to astrophysical plasmas. 

\begin{acknowledgments}
S. Dutta and S. Sengupta would like to acknowledge very fruitful discussions with our dear friend and colleague, Dr. R. Srinivasan, who unfortunately passed away during the preparation of this manuscript. 
\end{acknowledgments}

%\section*{Data availability}
%The data that supports the findings of this study are available within the article

\section*{References}
\nocite{*}
\bibliography{paper1}
\bibliographystyle{unsrt}
\end{document}